**Type of Presentation**:   **Topic:**  Speech Signal Processing

Oral: X          In-person: ✗

# How to Leverage DNN-based speech enhancement for multi-channel speaker verification?

**Sandipana Dowerah [1], Romain Serizel [1], Denis Jouvet [1], M.Mohammadamini [2], Driss Matrouf [2]**
[1] Université de Lorraine, CNRS, Inria-Loria., Nancy, France
[2] Avignon Université, Laboratoire Informatique d'Avignon, Avignon, France
E-mail: name.surname@loria.fr, name.surname@univ-avignon.fr

**Summary:** Speaker verification (SV) suffers from unsatisfactory performance in far-field scenarios due to environmental noise and the adverse impact of room reverberation. This work presents a benchmark of multichannel speech enhancement for far-field speaker verification. One approach is a deep neural network-based, and the other is a combination of deep neural network and signal processing. We integrated a DNN architecture with signal processing techniques to carry out various experiments. Our approach is compared to the existing state-of-the-art approaches. We examine the importance of enrollment in pre-processing, which has been largely overlooked in previous studies. Experimental evaluation shows that pre-processing can improve the SV performance as long as the enrollment files are processed similarly to the test data and that test and enrollment occur within similar SNR ranges. Considerable improvement is obtained on the generated and all the noise conditions of the VOiCES dataset.

## 1. Introduction

Speaker verification (SV) authenticates a person's identity based on his/her voice characteristics. Despite significant improvements in deep learning-based SV in close-talk or controlled scenarios, SV still suffers from unsatisfactory performance in far-field/ distant scenarios. Speech signals propagating over long distances are subject to fading, absorption, and reflection by various objects, which changes the pressure level at different frequencies and degrades speech quality. In real cases, these acoustic perturbations make far-field SV a challenging task. Several challenges have been organized over the past few years to address this problem, such as VOiCES from a distance challenge [12], Interspeech far-field speaker verification challenge [13], etc. The current state-of-the-art x-vector-based [8] approaches improved the SV performance significantly. But, these SV systems still suffer from severe performance degradation in noisy-reverberant scenarios that are typical of hands-free applications.

Speech enhancement can be used to improve the perceptual quality of speech by estimating clean speech signals for signals impacted by acoustic noise and reverberation. Besides denoising autoencoder [14], [15], neural beamforming [16], and dereverberation [17] have been extensively used as front-end processing of speech recognition [16, 18, 19]. But, only a few studies have examined the effectiveness of integrating beamforming and dereverberation with multichannel signals for SV in a noisy-reverberant environment [11, 20]. Mosner et al. employed mask-based beamforming combined with WPE to minimize the reverberation effect, but they studied only the reverberation effect, whereas reverberation and noise occur simultaneously in real scenarios [11]. Yang et al. jointly optimized neural networks that supported minimum variance distortionless response (MVDR) beamforming with WPE using a deep speaker embedding model [21]. Taherain et al. used an MVDR beamformer with Rank-1 approximation to search for the optimal beamformer from the variants of ideal ratio mask-based MVDR and generalized eigenvalue (GEV) beamformers [11]. Although often used in a multichannel context, most of these studies use single-channel data as an input to DNN, use matched train/test data, and report poor performance on real data. Moreover, prior works mainly used mask-based beamformers (MVDR or GEV) in the frequency domain, which typically degrades in causal and online scenarios [1] as frequency domain methods lack the reasonable size of frequency resolution and input signal length required for perceivable system latency.

This paper studies the benchmark speech enhancement as a multichannel pre-processing to SV in adverse acoustic conditions where noise and room reverberation distorts the target speech signal. We consider either filtering based on a deep neural network (DNN) or combining DNN and signal processing approaches. The DNN-based approach implements FaSnet [1], a state-of-the-art neural beamforming technique for speech enhancement. The second approach is an integration of FaSnet with Rank 1 multi-channel Wiener filter [2] and a de-reverberation algorithm [3]. We compared our pre-processing approach to the popular state-of-the-art pre-processing approaches from [11]. Furthermore, this work studies the impact of both approaches in different noisy and reverberated acoustic scenarios using various signal-to-noise (SNR) ratios for a multi-channel input signal. We also study the impact of data mismatch, robustness in low SNR scenarios, and generalization to unseen real recorded data. Additionally, we investigate the influence of quality (in terms of source to distortion and source to interference ratio) of the enhanced signals, which could be helpful in fine-tuning the front end of an SV system.

## 2. Use Case

The main purpose of this work is to solve the various challenges faced by a mobile security robot in the context of an SV. The performance of SV reduces drastically due to ambient noise leading to low SNR, internal robot noises leading to reverberation further degrading the SNR, and overlapping background speech. The main focus is on developing a multichannel speech enhancement as a pre-processing to the SV in the context of mobile security robot to identify a person on industrial premises during the inactivity period.

## 3. Problem Formulation

### 3.1. Signal Model

Considering the mixture of dry speech and noise as recorded by *K* microphones can be formulated with the short-time Fourier transform STFT as *y(T-F) = s(T-F) + h(T-F) + n(T-F)*, where *y(T-F), s(T-F), h(T-F)* and *n(T-F)* represent the STFT vectors of the noisy speech, dry speech, reverberated speech, and noise.

## 4. Multichannel Speech Enhancement

This section explains the integrated multichannel speech enhancement approach we developed for far-filed SV.

### 4.1. FaSNet

FaSnet (filter-and-sum network) is a filter-based beamforming approach suitable for real-time low-latency applications [1]. FaSnet incorporates a two-staged architecture. A beamforming filter for a chosen reference channel is computed in the first stage. The reference channel is randomly selected. The second stage uses the output filter from the first stage to estimate the beamforming filters for the rest of the channels. The input for both stages includes the target channel as well as the output of the normalized cross-correlation between channels as an inter-channel feature. Both stages use the temporal convolutional networks (TCN), enabling the lower latency processing of the FaSnet model. The training objective of the FaSnet model is to select a signal-level loss criterion based on the actual task needed to be solved.

We give noisy multichannel signals as input to FaSnet to separate noise and speech. We used FaSnet to obtain a first estimate of the speech signal s(T-F) and the noise signal n(T-F). These estimates are then used to compute the T-F masks:

$$M_s(T-F) = \frac{|S(T-F)|}{|S(T-F) + \max(|n(T-F)|, \varepsilon)|} \quad (1)$$

$$M_n(T-F) = \frac{|n(T-F)|}{|S(T-F) + \max(|n(T-F)|, \varepsilon)|} \quad (2)$$

where, $\varepsilon$ is 1 x 10$^{-16}$.

### 4.2. Rank-1 MWF

MWF is designed to minimize the mean squared error (MSE) criterion between the record mixture and the target speech.

$$J(w) = E\{|s_1 - w^H y|^2\} \quad (3)$$

where $s_1$ is the clean speech signal from the first channel, $E$ is the expectation operator, and $.^H$ denotes the Hermitian transpose. The filter $w$ that minimizes the MSE criterion [equation number] is the MWF that can be expressed as below;

$$\widehat{W}_{MWF}(f) = R_{ss}(f) + R_{nn}(f)^{-1} R_{ss}(f) u_1 \quad (4)$$

Where $R_{ss}$, $R_{nn}(f)$ are spatial correlation matrix for the speech and noise, respectively and $u_1 = [1,..., 0]^T$.

It is possible to introduce a trade-off parameter µ which controls the tradeoff between the interference reduction and the desired signal distortion [23]. We then obtained the speech distortion weighted (SDW) MWF that can be expressed as;

$$\widehat{W}_{SDW-MWF}(f) = R_{ss}(f) + \mu R_{nn}(f)^{-1} R_{ss}(f) u_1 \quad (5)$$

If the desired signal comes from a single source, the speech correlation matrix $R_{ss}$ is theoretically of Rank-1. Forcing this matrix to its Rank-1 approximation leads to the so-called Rank-1 version of the filters described above. In the remainder of the paper, we use the Rank-1 approximation of the SDW-MWF.

The computation of MWF requires the estimation of the speech and noise correlation matrices. The estimated T-F masks of speech and noise are used to compute the spatial correlation matrices $R_{ss}(f)$ and $R_{nn}(f)$ that are needed to derive the MWF. The correlation matrices are obtained as;

$$R_{ss}(f) = \frac{1}{T} \sum_{t=0}^{T-1} \widehat{s}(T-F) \widehat{s}(T-F)^H \quad (6)$$

Note that the noise correlation matrix can be obtained similarly as in Eq.[6].

### 4.3. Weighted Prediction Error (WPE)

WPE is used for alleviating degradation performance in speech recognition, mostly in the case of a far-field scenario. The de-reverberated signal is obtained by subtracting the filtered signal from the observed signal denoted by;

$$d(\hat{s}) = \hat{s}(t) - \sum_{k=1}^{N} \hat{w}(k)\, h(T - k)$$

Where, $\hat{s}$ is reverberated signal at time t and $d(\hat{s})$ is de-reverberated signal using WPE algorithm. $\hat{s}$ denotes the $k^{th}$ tap of the N-taps. WPE filter is $W = [W_1, ..., W_N]^T$.

## 5. Datasets

### 5.1 Synthetic Dataset

We generated a synthetic dataset, namely, RoboVoices simulating real room environments with additive noise and reverberation from dry speech segments. Designing such a dataset is necessary as training speech enhancement approaches require ground-truth knowledge about the target speech and, to some extent, the degradation. This information is not available in the available corpora for far-field SV.

### 5.1.1. Speech Data

We use the dry speech data from the clean subset of the Librispeech [4] corpus, which is approximately 1000 hours of English speech data collected as part of the Librivox project. We randomly selected around 10000 files from the dry training subset of Librispeech and truncated them to 10 seconds duration for the training set, contributing to 25 hours of speech data.

For the evaluation of the SV system, we use the Fabiole speech corpus [5]. Fabiole is a French speech corpus consisting of around 6882 audio files from 130 native French speakers. The minimum duration of the speech file is 1 second, and the maximum is 46 seconds. The speech data of the corpus is collected from different French radio and television shows. For creating each evaluation set, we have used 1200 speech files from Fabiole representing 2 hrs of evaluation material.

### 5.1.2. Noise Data

We have collected realistic office noise from the Freesound platform [6]. The selected noise categories include door, keyboard, office, phone, background noise in the room, printer, fan, door knock, babble, environmental noise, etc. We split the dataset into a training set of 3725 clips and an evaluation set of 1000 clips.

We also evaluate our system's performance using MUSAN noise from the OpenSRL dataset [9]. We convolved the dry speech from Librispeech and noise from Musan with simulated RIR for training. The evaluation protocol is the same as RoboVoices except for the noise samples. The noise categories include dial tones, raindrops, etc.

### 5.1.3. Room Impulse Response

To simulate room effects, we have generated an RIR corpus of 10000 rooms for training and 3600 for evaluation with the pyroomacoustics toolbox [7]. For training, the room length was chosen between 3-8 m, the width was chosen between 3-5 m, and the height was chosen between 2-3 m. The absorption coefficient was drawn randomly such that the room's RT-60 was between [200-600] ms. The minimum distance between a source and the wall is 1.5 m, and 1 m between the wall and the microphones. The RIR for the evaluation set was generated with the same room dimension as in the training set, but the absorption coefficient was selected to obtain an RT-60 of 400 ms.

The final RoboVoices corpus for training and evaluation is created by first convolving the dry speech and noise with the simulated RIRs. We then added the convolved dry speech and convolved noise to obtain the noisy signal. We randomly select the noise samples from Freesound and the dry speech from Librispeech for the training set. The SNR is drawn randomly with a uniform distribution between 0-10 dB. For the evaluation set, the generation process is similar, except that we draw the SNR values in 5, 10, and 20 dB, and the process is applied to each speech segment from the Fabiole dataset. In total, we have generated 10000 mixtures for training and 3600 mixture for evaluation.

### 5.2. VOiCES

We evaluate our approach to the VOiCES challenge 2019 dataset [12]. Among 11 microphone positions in the Eval set, we select three representative positions: 2, 4, and 9. We select the signal from these three microphones confirming all three are in mid-distance from the speaker and are close to building a "virtual" microphone antenna.

## 6. Experimentation

### 6.1. Experimental Set-up

The speech and noise signals are sampled at 16 kHz. We provide multichannel speech signal as input to FaSNet with a 4 ms window size and context size of 16 ms. We trained the FaSNet model with SDR loss and SI-SNR (scale-invariant source-to-noise ratio) loss [24]. We employed the dual-path RNN (DPRNN) with an encoder dimension of 50, a chunk size of 50, and a hopping window of 35 dimensions. We use the source-separated outputs from the FaSNet model to compute the target masks. The FaSNet implementation is used from the Asteroid toolbox [25] and replaced the TCN blocks with DPRNN in contrast to the original FaSNet architecture, where TCN is used to predict the beamformed filters.

The SDW-MWF operates on the T-F representation of the signal. STFT is computed with a window length of 512 samples, a hop size of 256 samples, and a Hann window. A single SDW-MWF is estimated for each speech clip. According to previous experiments, we set the μ parameter of the SDW-MWF to 0.1 to limit the amount of distortion introduced by the filter. We use WPE with the following parameters: 10 filter taps, a delay of 3 frames, 5 iterations of WPE, and an alpha of 0.9999.

*6.1.2. Speaker Verification*

Our SV is an x-vector-based system. The network is trained with data augmentation using different portions of Musan corpus (music, babble, noise, reverberation) [9] with 1 million augmented files from Voxceleb [26] and all the original files from Voxceleb 1 and 2 [10]. We use the Fabiole corpus for tests and enrollment. For enrollment, 3441 files are used, and the remaining files are used for the test. As input to the x-vector network, we extract Mel-frequency cepstral coefficients normalized by Cepstral Mean-Variance Normalization. We removed the non-speech frames with a voice activity detector. The Probabilistic Linear Discriminant Analysis (PLDA) classifier used for scoring is trained on 200k x-vectors extracted from Voxceleb. Before training the PLDA, x-vectors are centered, and their dimensionality is reduced to 128 with linear discriminant analysis. The PLDA scoring system is retrained on the enrollment set. Kaldi[1] is used to process all the steps of SV.

*6.2. Evaluation*

The SV system is evaluated using an equal error rate (EER). The bootstrap algorithm presents all metrics with a 95% confidence interval [27]. We compute EER on dry speech and reverberated speech (as a reference point), the input mixture, and the signals estimated with different speech enhancement algorithms.

## 7. Results and Analysis

*Table-1*: *EER (%) on RoboVoices using different pre-processing methods. The confidence interval is 0.1.*

| Pre-processing/SNR | 5 | 10 | 20 |
|---|---|---|---|
| *Unprocessed* | 34.4 | 28.0 | 22.2 |
| *FaSnet [1]* | 45.7 | 39.0 | 31.5 |
| *BLSTM MVDR Rank1 [11]* | 32.3 | 26.6 | 22.1 |
| *BLSTM GEV-BAN [11]* | 32.5 | 26.8 | 21.9 |
| *FaSnet Rank1 MWF WPE* | **27.1** | **23.2** | **19.7** |

Table-1 shows the results of our experiments using different pre-processing approaches on the RoboVoices dataset on various SNR conditions. We implement the BLSTM-based approaches from [11] and consider them as the baseline. FaSnet degrades the SV performance as FaSnet introduces artifacts that distort the signal quality. We observed that integrating FaSnet with signal processing methods consistently improves against the baseline systems BLSTM MVDR Rank-1 and BLSTM GEV-BAN. Error reduction in multi-channel SV is greater, especially in low SNR, with a 7% reduction at 5 dB showing robustness to low SNR conditions for an unprocessed noisy-reverberated signal. Thus, supporting the argument that Rank-1 MWF is robust to low SNR scenarios.

*Table-2:* *EER (%) on different noise conditions of the VOiCES Eval dataset. The confidence interval is 0.2.*

| | Clean | Babble | TV | Music |
|---|---|---|---|---|
| *Unprocessed* | 4.4 | 9.2 | 7.9 | 8.4 |
| *FaSnet* | 4.4 | 7.8 | 7.4 | 7.9 |
| *MVDR Rank1* | 4.3 | 7.3 | 6.5 | 6.9 |
| *FaSnet Rank1 MWF WPE* | **4.0** | **6.3** | **6.0** | **6.4** |

Table-2 presents the results obtained on the publicly available VOiCES Eval dataset [12] for various distractor noise conditions. We selected the microphone which was closest to the speaker as a reference microphone. As expected, the condition with no noise distractor (Clean in Table 2) resulted in the best performance across the approaches. The baseline BLSTM-based approaches perform poorly compared to the FaSnet-based approaches in all the noise conditions. With an EER of 9.2% without any pre-processing, Babble seems to be the most challenging condition due to overlapping speech interference as well as its similarity to the desired clean speech. The proposed system improved the performance of Babble with an EER of 6.3 %. Furthermore, FaSnet Rank-1 MWF WPE achieves the best performance across the noise conditions, demonstrating our approach's efficacy even though the model was trained on synthetic data generated for generic, possibly mismatched, and spatial scenarios. We have also experimented with enrollment in match pre-processing conditions showing its impact in SV.

Table-3 reports the performance on the RoboVoices dataset for different pre-processing conditions and depending on the enrollment condition. Performing the enrollment and test with matched acoustic conditions alleviates the effect of reverberation. but this is hardly the case for additive noise. Pre-processing consistently improves the SV performance, but the effectiveness is more evident when the enrollment is done in matched pre-processing conditions (diagonal). FaS Rank-1 MWF WPE obtained the best EER performance for a noisy and reverberated input over the baseline approach.

---
[1] https://github.com/kaldi-asr/kaldi

**Table-3:** *EER (%) on matched pre-processing conditions on the RoboVoices dataset. We processed both enrollment and test data using the same range of SNR. The average confidence interval is 0.1.*

| Test data | Dry speech | Reverb. Speech | Noisy | BLSTM MVDR Rank-1 | FaSnet Rank-1 MWF WPE |
|---|---|---|---|---|---|
| Dry speech | **14.9** | 15.4 | 16.7 | 16.1 | 15.7 |
| Reverb. speech | 20.6 | **19.8** | 20.5 | 20.4 | 20.1 |
| Noisy | 28.2 | 24.9 | **23.8** | 24,9 | 24,3 |
| BLSTM MVDR Rank -1 | 27.0 | 24.2 | 23.4 | **21.3** | 22.5 |
| FaSnet Rank-1 MWF WPE | 23.3 | 22.8 | 21.5 | 22.4 | **19.2** |

## 8. Conclusion

This work presents the benchmark speech enhancement pre-processing approach to multi-channel speaker verification in a far-field noisy-reverberated environment. We experimented with both DNN and a combination of DNN with signal processing methods as a front end to the state-of-the-art x-vector speaker verification system. Experimental evaluations on synthetic and VOiCES datasets show that combining DNN with signal processing methods significantly improves speaker verification performance. Moreover, the combined DNN and signal processing approach show more robustness to low SNR scenarios. Additionally, experimentation with enrollment shows that performing the test and enrollment with matched acoustic conditions alleviates the effect of reverberation. Our approach demonstrated the best performance across the noise conditions on the VOiCES dataset even though the model was trained on synthetic data. This shows that our approach generalizes to unseen real recorded data.

## 9. Acknowledgements

French National Research Agency supports this work for the ROBOVOX project (ANR-18-CE33-0014)). Experiments were partially carried out using the Grid5000 testbed supported by a scientific group of Inria, including CNRS, RENATER, and other Universities and organizations hosted by the University of Lorraine.

---

[2] https://www.grid5000